\documentclass[amsmath,showpacs,twocolumn,aps,prl]{revtex4}
\usepackage{braket}
\usepackage{amsmath,amsfonts}
\usepackage{graphicx}
\usepackage{color}
\usepackage{enumerate}
\usepackage{hyperref}
\makeatletter

\newcommand{\Rmnum}[1]{\expandafter\@slowromancap\romannumeral #1@}
\makeatother
\newcommand{\be}{\begin{equation}}
\newcommand{\ee}{\end{equation}}
\newcommand{\bea}{\begin{eqnarray}}
\newcommand{\eea}{\end{eqnarray}}

\date{\today}

\begin{document}

\title{Nonlinear Optical Conductivity in Graphene and other 2-Band 2-D Materials}

\author{Sushanta Dattagupta$^{1,}$\footnote{sushantad@gmail.com}, and Manvendra Singh$^{2,}$\footnote{manvendra16@iiserb.ac.in}}
\affiliation{\mbox{$^1$}{Bose Institute, Kolkata 700054, India}\\
\mbox{$^2$}{Department of Physics, Indian Institute of Science Education and Research, Bhopal, India}\\
}

\begin{abstract}
Graphene, Silicene, $\mathrm{MoS}_2$ and other similar two-dimensional structures have unusual electronic properties that lend themselves to exotic device applications. These properties emanate from the fact that the electrons are endowed with Dirac fermion-like attributes. Thus these materials are not only characterized by certain fundamental principles, they also have amazing practical uses. Our emphasis here is on one such basic property concerning nonlinear response to time-dependent electric fields. We set up a first principle quantum master equation for the underlying density operator which is based on microscopic interactions between the Dirac electron with phonons and other electrons. While such an equation has general applicability to a variety of non-equilibrium phenomena in two-band systems, we focus onto the case of nonlinear optical conductivity. The derived results are separately analyzed for graphene, silicene and $\mathrm{MoS}_2$, and comparison made with other known results.  
\end{abstract}


\maketitle

\section{1. Introduction}
Two dimensional electronic materials such as graphene, silicene, $\mathrm{MoS_2}$ etc., notwithstanding their myriad technological and device applications, are a gold mine for testing theoretical concepts of great contemporary interest~\cite{Geim1,Geim2,katsnelson,Houssa}. These involve novel ideas of  the chemistry of hybridized carbon orbitals~\cite{saito}, ultra high mobility~\cite{bolotin,du}, spin-orbit interaction~\cite{ezawa}, Andreev reflection and Klein tunneling~\cite{beenakker1, beenakker2, stander}, magneto-resistance and weak localization~\cite{falko,tikhonenko,chen}, quantum Hall effect~\cite{kim,novoselov,du2,bolotin2}, spintronics~\cite{han,inoue}, valleytronics~\cite{mak} and so on. Much of the extraordinary properties of these materials manifest themselves on account of their electrons behaving as Dirac Fermions in the low energy physics sector. Indeed, all the interesting physics emanates from the underlying Dirac equation, though the electrons travel with not the speed of light but much smaller Fermi velocity. Thus these materials have turned out to be a laboratory for the realization of relativistic quantum mechanics, albeit in a non-relativistic setting of solid state physics~\cite{katsnelson}.\\

Given this background to all the extraordinary phenomena in graphene, and graphene-like materials, one aspect that is going to particularly occupy our attention in the present study is the breakdown of linear response theory in connection to a transport property such as the frequency-dependent dynamic conductivity~\cite{mishchenko}.  The necessity to go beyond the usual `Kubo regime' hinges on the fact that our interest is in the systems for which  the Fermi energy $\epsilon_F$ is vanishingly small. In our calculation of the nonlinear electrical conductivity we follow a density matrix approach for a general two band model~\cite{mishchenko,ishikawa, agarwal}. But, in a departure from the existing treatments in which dissipative terms are incorporated phenomenologically, we employ a first-principles approach wherein these terms occur in a natural manner from the considerations of probability conservation and detailed-balance. Further, our approach is couched in the framework of quantum dissipation of two-state systems in contact with phonon and electronic baths~\cite{weiss11, SDSP}. Though the quantum baths are not treated in detail here, the indicated method opens up possibilities for an elaborate analysis of dissipation in graphene and graphene-like systems.

An outline of the paper is as follows. In section 2, we introduce the relevant Hamiltonian for graphene, silicene and $\mathrm{MoS}_2$ within the product Hilbert space of two distinct spin `variables' – one operating within the two sub-lattice states of the material, and the other within the actual spin-states of the electron. The form of the Hamiltonian is illustrated by citing three different materials of grapheme, silicene and $\mathrm{MoS}_2$, in sub-sections 2a, 2b and 2c, respectively. Section 3 deals with the von-Neuman – Liouville equation for the density operator. That equation has two parts, one governed by systematic Hamiltonian dynamics in the presence of an external electric field, and the other concerned with dissipative dynamics triggered by the surrounding heat bath. These issues are separately treated in sections  3a and 3b respectively. While discussing the systematic dynamics in Section 3a, we employ a `Rotating Wave Approximation' (RWA) in which rapidly oscillating terms that have no influence in the non-transient regime, are dropped. Section 3b is devoted to standard rate equation method in which the need for probability conservation and detailed-balance automatically emerge from microscopic considerations. While section 3 contains a general framework for dealing with dissipative effects in the 2-d materials at hand, we specialize in section 4 to the calculation of nonlinear optical conductivity. Once again, separate results are presented for the three systems of graphene, silicene and $\mathrm{MoS}_2$. These results are analyzed in section 5 as a prelude to offering certain concluding remarks in section 6 .\\

\section{2. The Dirac Hamiltonian}

The generic Hamiltonian of our focus is an explicitly time dependent one given by ~\cite{mishchenko},
\begin{equation}
\mathcal{H}_{t} = \mathcal{H}_0({\sigma}, {s}) + \mathcal{H}_{I}({\sigma},s) \, \sin({\omega t}) \label{eq:AB1}
\end{equation}
where $ \mathcal{H}_{0}({\sigma},{s)}$ is an unperturbed Hamiltonian that lives in the product space of the two distinct dynamical variables $\sigma$- a pseudo spin and $s$- the real spin of the electron. The time $t$ is inserted as a suffix to underscore the fact that the underlying time dependence is only a parametric (and not a dynamic) one. The meaning of $\sigma$ and $s$ is clarified in the context of three different electronic materials, below. The time dependent interaction term arises from an externally applied electronic field of amplitude $E$ and frequency $\omega$. Evidently, $\mathcal{H_{I}}$ being off-diagonal in the representation in which $H_0$ is diagonal, would cause quantum transitions amongst the eigenstates of $H_0$. These transitions would imply gain or loss in energy of the unperturbed system which would have to be compensated by corresponding transitions in the surrounding heat bath (treated in section 3 below), thus leading to dissipation. One measurable quantity that characterizes dissipation is a non-linear electrical conductivity which has to be calculated to all orders in $E$ and $\omega$. The concrete case study of Eq.~(\ref{eq:AB1}) can be effected in the context of three different materials, as described in the following sub-sections.

\subsection{2a. Graphene}
\begin{figure}
	\centering
	\includegraphics[width=0.5\textwidth]{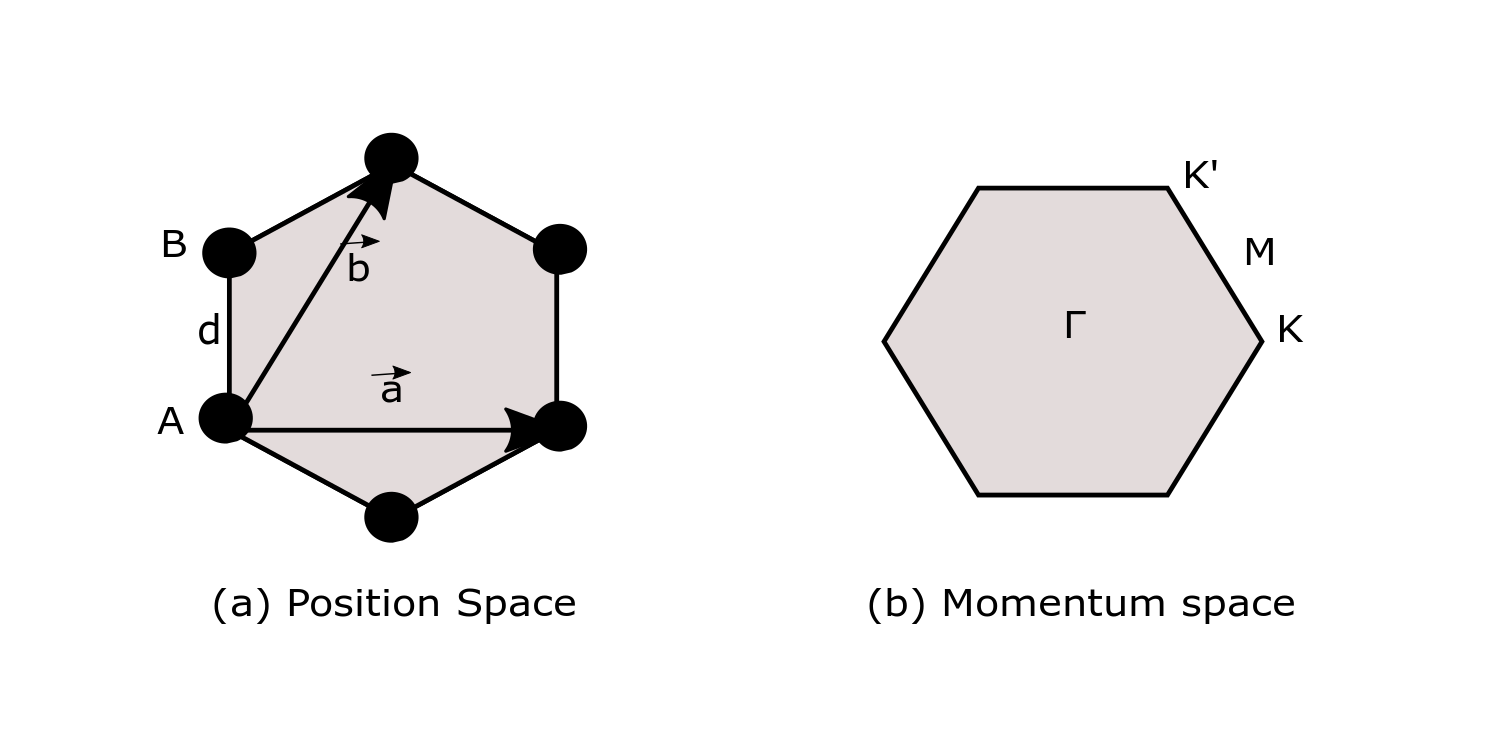}
	\caption{\label{fig:Graphene} Position space and momentum space picture for graphene}
\end{figure}

 As our first example, we consider graphene which has a 2-dimensional honeycomb structure as shown in Fig.~(\ref{fig:Graphene}), It is evident that there are two inter-penetrating triangular lattices marked A and B. If only nearest neighbour tunneling is allowed, an electron can only jump from an A-site to a B-site and vice-versa. The corresponding tight-binding Hamiltonian may be written as 
\begin{equation}
\mathcal{H}_{0}= - J \sum_{(i{A},j{B})}({\ket{\uparrow}}_{i} {\bra{\downarrow}}_{j} + {\ket{\downarrow}}_{i} {\bra{\uparrow}}_{j}),
\end{equation}
where $J$ is the (constant) tunneling (or hopping) energy, $\ket{\uparrow} ( \ket{\downarrow})$ denotes presence (absence) at the site A (B). The above can be rewritten in the Bloch-Fourier space as
\begin{equation}
\mathcal{H}_{0} = -\sum_{k} [J(k) \ket{\uparrow, k} \bra{\downarrow, k} + J^{*}(k) \ket{\downarrow, k} \bra{\uparrow, k}  ].
\end{equation}
Here the wave-vector dependent tunneling $J (k)$ is complex that assumes a linear form near the $K $ and $ K'$-points of the Brillouin zone in the reciprocal space (Fig.  ~\ref{fig:Graphene}). Thus the Hamiltonian for a given $k (=p,\, \hbar=1)$ has the following $2\times 2$ representation in the space of the pseudo-spin $\sigma$
\begin{equation}
\mathcal{H}_{0} = v_{F} \, (\vec{\sigma} .\vec{k}) = v_{F} \, k(\tau \sigma_x \, \cos\chi_k+ \sigma_y \, \sin\chi_k) ,\label{eq:AB2}
\end{equation}
where $\ket{\uparrow}$ and $\ket{\downarrow}$ are the eigenstates of $\sigma_z$ with eigenvalues + 1 and -1 respectively.
 In the above equation,  $\tau=\pm 1$ for $K / K'$ valleys, $\chi_k$ is the angle made by the 2-dimensional vector $\vec{k}$ with the $X$-axis and $v_F(=3 J d/{2\hbar})$ is the Fermi velocity that is estimated to be of the order of $10^{6} \text{m/sec}$, $d$ being the lattice parameter (Fig.~\ref{fig:Graphene}) . Comparing Eq.~(\ref{eq:AB2}) with Eq.~(\ref{eq:AB1}) we note that $\mathcal{H}_{0}$ is independent of the spin degree of freedom s, for graphene. Thus the eigenstates of $\mathcal{H}_0$ are the electronic band states associated with the conduction band ($\ket{c_k}$) and the valence band ($\ket{v_k}$), thus
\begin{equation}
\ket{c_{k\tau}}=
 \frac{1}{\sqrt{2}}\begin{bmatrix}
\tau e^{-i \tau \chi_k/2} \\
e^{i\tau \chi_k /2 } 
\end{bmatrix}
 \label{eq:gcket}
 \end{equation}
and 
\begin{equation}
\ket{v_{k\tau}}=
 \frac{1}{\sqrt{2}}\begin{bmatrix}
\tau e^{-i \tau \chi_k/2} \\
-e^{i\tau \chi_k /2 } .
\end{bmatrix}
\label{eq:gvket}
 \end{equation}
The equations~(\ref{eq:gcket}) and (\ref{eq:gvket}) constitute our generic two state system(for fixed momentum, valley index and spin index). It is the linearity in the momentum (or equivalently the wave vector) of the Hamiltonian that justifies the epithet: ‘massless Dirac Fermions’ for the electrons in graphene, as stated earlier.
Following~\cite{mishchenko} we now add to Eq.~(\ref{eq:AB2}) a term associated with a sinusoidal electric field of amplitude $E$ and frequency $\omega$ in the x-direction, 
\begin{equation}
\mathcal{H}_t = v_{F} (\vec{\sigma}.\vec{k}) + \tau \sigma_{x} \, (e \, v_{F} E/\omega)\, \sin(\omega t), \label{eq:I7}
\end{equation}
$e$ being the electron charge. Comparing the above equation with Eq.~(\ref{eq:AB1}), we have 
\begin{equation}
\mathcal{H}_I=\tau \frac{e v_F E}{\omega} \sigma_x .
\label{eq:HI}
\end{equation}

\subsection{2b. Silicene} 
\begin{figure}
	\centering
	\includegraphics[width=0.3\textwidth]{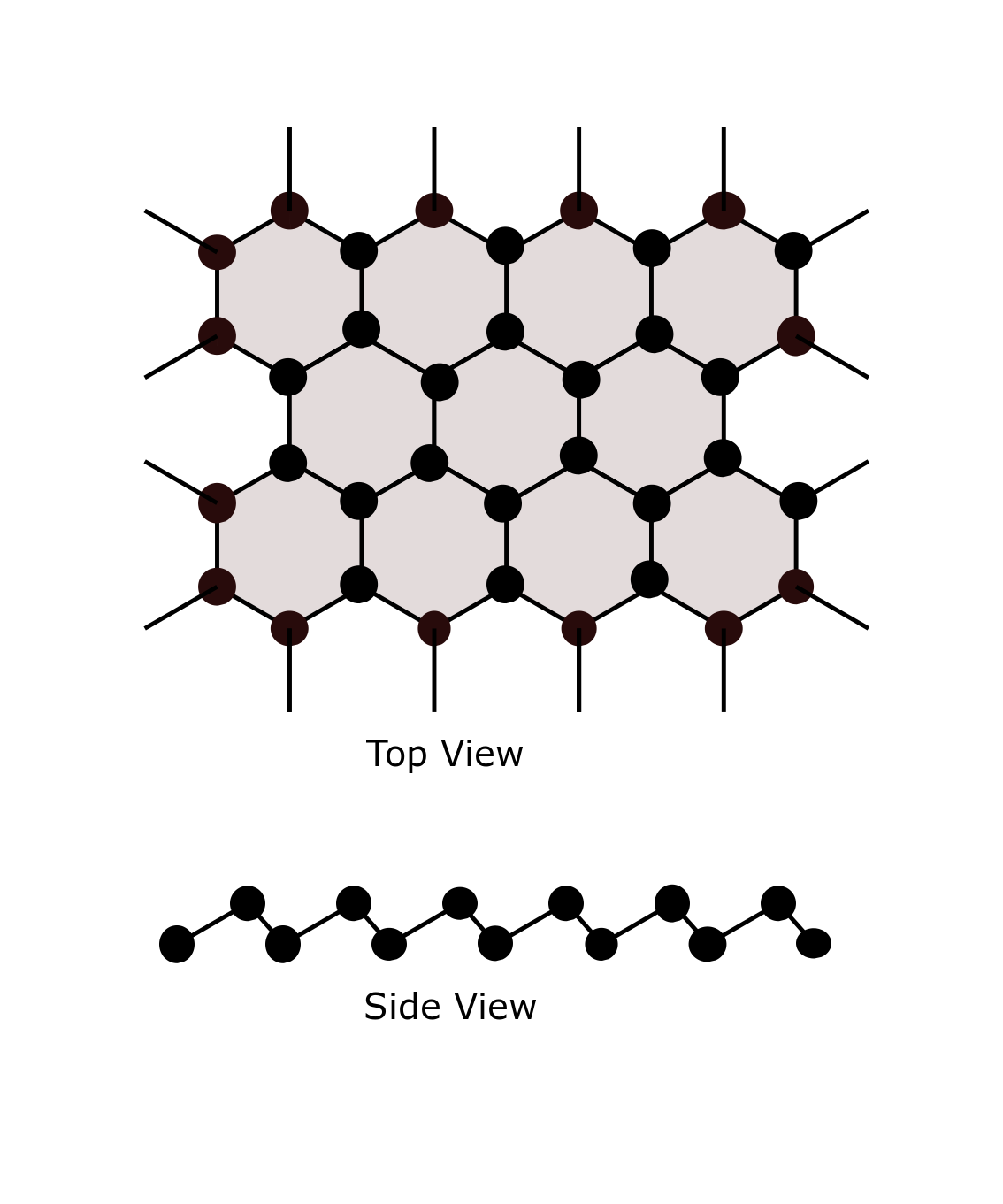}
	\caption{ Top view and side view of Silicene}\label{fig:Silicene}
\end{figure}

Our next example is that of  silicene which is made up of a single layer of silicon  atoms  forming a   buckled honeycomb structure  (see Fig.~\ref{fig:Silicene}).
\begin{figure}
	\centering
	\includegraphics[width=0.3\textwidth]{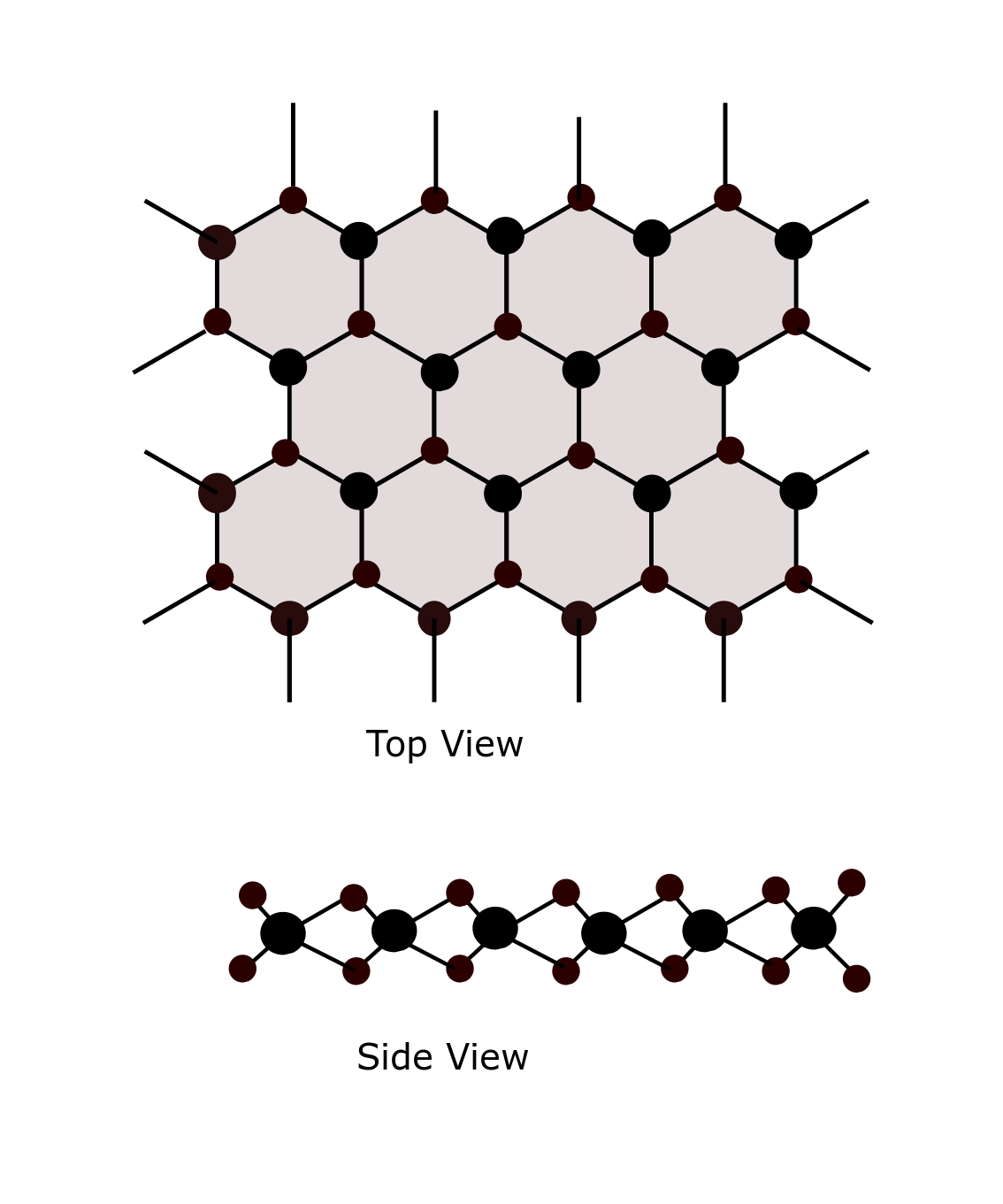}
	\caption{\label{fig:MoS2} Top view and side view of $\mathrm{MoS_2}$}
\end{figure}
The low energy Hamiltonian  about the $K/K'$ points has the following form~\cite{ezawa},
\begin{equation}
\mathcal{H}_0 =  v_{F} (\tau k_x \sigma_x -k_y \sigma_y)  -\tau \lambda_{SO} s_z\sigma_z   + l E_z\sigma_z, \label{eq:AB3}
\end{equation}
where $\tau=\pm 1$ again represent $K/K'$ valleys  and  $\sigma$- and $s$- denote  Pauli matrices corresponding to the sub-lattice and spin degrees of freedom, respectively. The first term of Eq.~(\ref{eq:AB3}) is the linear Dirac term while the second term is due to the  spin-orbit interaction (with coupling constant $\lambda_{SO}$) which is responsible for opening up a gap in the spectrum. The third term of the Hamiltonian serves as a gap tuning parameter which arises when an external electric field $E_z$ is applied perpendicular to the surface, thus generating different potentials on the A and B lattice sites. 
 Note that the only relevant component of the electronic spin which is to be reckoned with lies along the $z$-axis. However, as $s_z$ is a good quantum number, the static Hamiltonian simply splits into two disconnected sectors associated with spin-up and spin-down states (with $s_z\rightarrow \pm 1$), thus 
 \begin{equation}
\mathcal{H}_0^{s} =  v_{F} (\tau k_x \sigma_x -k_y \sigma_y)  -s \tau \lambda_{SO} \sigma_z   + l E_z\sigma_z. \label{eq:AB4}
\end{equation}
The eigenvalues are $\epsilon_{s\tau}= \pm \sqrt{v_F^2k^2 +\Delta_{s\tau}^2} $, where $\Delta_{s\tau}=  l E_z -s \tau \lambda_{SO}     $. The corresponding eigenstates are, 
\begin{equation}
\ket{c_{s\tau}}=\frac{1}{\sqrt{2 |\epsilon_{s\tau}| \eta_{s\tau} }}\begin{bmatrix}
\tau k v_F e^{i \tau \chi_k} \\
 \eta_{s\tau}
\end{bmatrix} ,
\label{eq:scket}
\end{equation}
and 
\begin{equation}
 \ket{v_{s\tau}}=\frac{1}{\sqrt{2 |\epsilon_{s\tau}| \bar{\eta}_{s\tau} }}\begin{bmatrix}
\tau k v_F e^{i \tau \chi_k} \\
 -\bar{\eta}_{s\tau}
\end{bmatrix},
 \label{eq:svket}
\end{equation}
where $\eta_{s\tau}=  \sqrt{v_F^2k^2 +\Delta_{s\tau}^2}  -  \Delta_{s\tau} $ and $\bar{\eta}_{s\tau}=  \sqrt{v_F^2k^2 +\Delta_{s\tau}^2}  +  \Delta_{s\tau} $. 
The perturbation $\mathcal{H}_{I}$ (i.e. the second term in Eq.~\ref{eq:I7}) is taken to be of same form, as the direct coupling of the electric field to the spin $s$ is small.

\subsection{2c. $\text{MoS}_{2}$}
The third example considered is a monolayer of transition-metal dichalcogenide  $\text{MoS}_{2}$ which is  composed of an 
$\text{Mo}$ atom caged-in  by six neighboring $\text{S}$ atoms that are arranged in a prismatic geometry. The top view of monolayer  $\text{MoS}_{2}$   has a hexagonal structure  with the alternate sites occupied by $\text{Mo}$ and $\text{S}$ atoms (see Fig.~\ref{fig:MoS2}).
Unlike graphene, the inversion symmetry is explicitly broken in the monolayer of
 $\text{MoS}_{2}$ making it amenable to valleytronics and valley-dependent optical excitations~\cite{yao1}. It also supports a strong spin-orbit interaction making it  an ideal 2D-system to study spintronics~\cite{min,yao2}.  

The low-energy unperturbed Hamiltonian about the $K/K'$ points is given by~\cite{yao3},
\begin{equation}
\mathcal{H}_0= v_F(\tau k_x \sigma_x  +  k_y \sigma_y  )   +\frac{\Delta}{2}\sigma_z  -\lambda \tau \frac{\sigma_z-1}{2}s_z, \label{eq:mos2}
\end{equation}
where, once again, $\tau=\pm 1$ represent $K/K'$ valley and  $\sigma$ and $s$ are Pauli matrices corresponding to the sub-lattice and spin degrees of freedom, respectively. Note that the last term in Eq.~(\ref{eq:mos2}) represents the spin-orbit interaction. As before, spin is a good quantum number and the Hamiltonian can be decoupled in to the $s=\pm 1$ 
sectors.  The eigenvalues are $\epsilon_{s\tau}= \pm \sqrt{v_F^2k^2 +\Delta_{s\tau}^2}  +\frac{\lambda \tau s}{2}$, where $\Delta_{s\tau}=\frac{\Delta -\lambda \tau s }{2} $. The corresponding eigenstates are, 
\begin{equation}
\ket{c_{s\tau}}=\frac{1}{\sqrt{2 |\epsilon_{s\tau}| \eta_{s\tau} }}\begin{bmatrix}
\tau k v_F e^{-i \tau \chi_k} \\
 \eta_{s\tau}
\end{bmatrix} ,
\label{eq:mcket}
\end{equation}
and 
\begin{equation}
 \ket{v_{s\tau}}=\frac{1}{\sqrt{2 |\epsilon_{s\tau}| \bar{\eta}_{s\tau} }}\begin{bmatrix}
\tau k v_F e^{-i \tau \chi_k} \\
 -\bar{\eta}_{s\tau}
\end{bmatrix},
 \label{eq:mvket}
\end{equation}
where $|{\epsilon}_{s\tau}| =   \sqrt{v_F^2k^2 +\Delta_{s\tau}^2} $, $\eta_{s\tau}=  \sqrt{v_F^2k^2 +\Delta_{s\tau}^2}  -  \Delta_{s\tau} $ and $\bar{\eta}_{s\tau}=  \sqrt{v_F^2k^2 +\Delta_{s\tau}^2}  +  \Delta_{s\tau} $. Once again, our goal will be to study the 
non-linear current response of $\text{MoS}_2$ subject to an oscillating electric field term in the Hamiltonian given by $\tau \sigma_x(ev_FE/w)\sin(\omega t )$.
\section{3. Quantum Master Equation for the Density Operator}
With the Hamiltonian (Eq.~(\ref{eq:AB1})) at hand, applicable to nonlinear electrical conductivity measurements in graphene, silicene and $\mathrm{MoS_2}$ etc., the Von-Neumann-Liouville density operator equation can be written as
\begin{equation}
\frac{\partial \rho}{\partial t} = - i [\mathcal{H}_{t}, \rho_{t} ] \,. \label{eq:AB5}
\end{equation}
In order to arrive at a master equation for the density operator one has to add to the Hamiltonian $\mathcal{H}_{t}$ in Eq.~(\ref{eq:AB1}), the coupling to a heat bath. The later, in the present instance, comprises phonon and electron interactions. Upon averaging over the bath in the spirit of Lindblad~\cite{weiss11}, the averaged density operator can be split as:
\begin{equation}
\frac{\partial \rho (t)}{\partial t}=\frac{\partial \rho_S (t)}{\partial t}+\frac{\partial \rho_D (t)}{\partial t} \,,
\label{eq:RhoEvoltuion}
\end{equation}
where the subscript S denotes systematic component whereas the subscript D signifies the dissipative part. Needless to say, the heat bath has no influence on the systematic part $\rho_{S}$. We deal with these two terms separately in the following two sub-sections.
\subsection{3a. Systematic Dynamics in RWA} 
Given that our natural basis states in the two-band mode are the conduction and valence states we prefer to diagonalize the unperturbed Hamiltonian in that representation. Further, in order to avoid clumsy notation we simply re-express the conduction and valence states by $\ket{c}$ and $\ket{v}$ respectively. Thus
\begin{equation}
\mathcal{H}_{0}=\left(\frac{\epsilon_c + \epsilon_v}{2}\right) I+\frac{\epsilon_c - \epsilon_v}{2}(\ket{c}\bra{c}-\ket{v}\bra{v}),
\end{equation}
wherein all the indices $k$, $s$ and $\tau$ have been subsumed for the sake of brevity. For instance, in graphene,
\begin{equation}
\mathcal{H}_0 = k v_F (\ket{c}\bra{c}-\ket{v}\bra{v}).
\end{equation}
We note from the discussion above that though the structure of $\mathcal{H}_0$ is different in the three distinct cases of graphene, silicone and $\mathrm{MoS}_2$ (because of the additional presence of the spin s) the form of $\mathcal{H}_I$ has been assumed to be the same in all the three cases.\\
As our strategy is to work in the representation in which $\mathcal{H}_0$ is diagonal (in the space of conduction and valence states of $\ket{c}$ and $\ket{v}$), it is natural for us to go to the `interaction picture' in which,
\begin{equation}
\tilde{\rho}_S(t)= e^{i \mathcal{H}_{0} t} \rho_S(t)  e^{-i \mathcal{H}_{0} t} \label{eq:T1}
\end{equation}
that yields,
\begin{equation}
\frac{\partial \tilde{\rho}_S}{\partial t} = - i \, [\tilde{\mathcal{H}}_{I}, \tilde{\rho}_S(t) ] \sin(\omega t),
\end{equation}
where,
\begin{equation}
\tilde{\mathcal{H}}_{I} = \tau\frac{e v_F E }{\omega} e^{i \mathcal{H}_{0} t} \sigma_x  e^{-i \mathcal{H}_{0} t} .
\end{equation}
Rewriting the above in the basis representation of $\ket{c}$ and $\ket{v}$, and ignoring all the rapidly oscillating terms such as $e^{\pm i\omega t}$, $e^{\pm i(\omega+\omega_{cv})t}$ (here, $\omega_{cv}=\epsilon^{0}_{c}-\epsilon^{0}_{v}$), we arrive at an `effective' interaction Hamiltonian:
\begin{equation}
\mathcal{H}_{eff.}(t)=  \tau\frac{e v_F E }{4\omega}[i~\bra{c}{\sigma_{x}}\ket{v} \ket{c}\bra{v} e^{-i(\omega - {\omega}_{cv})t}+h.c.]. 
\label{eq:AB9}
\end{equation}
Therefore, as in ~\cite{walls} we shall work with an effective Von-Neumann Liouville equation
\begin{equation}
\frac{\partial \tilde{\rho}_S(t)}{\partial t} = - i [\mathcal{H}_{eff.}(t), \tilde{\rho}_S(t) ], \label{eq:AB7}
\end{equation}
It is instructive to spell out what $\mathcal{H}_{eff.}$ is, say for graphene, using the form of $\mathcal{H}_{I}$ given by the second term of Eq.(7) and the fact that $\epsilon^{0}_{c}=k v_{F}$ and $\epsilon^{0}_{v}=-k v_{F}$. We find 
\begin{equation}
\mathcal{H}_{eff}(t)=\, \frac{\Omega_k}{2} \,\ket{c}\bra{v} e^{-i \Delta_k t}  + h.c.,
\end{equation}
where $\Delta_k = \omega - \omega_{cv}$ and $\Omega_k=(\frac{e v_F E}{2 \omega})\sin\chi_k$  is the `Rabi frequency'.

Using Eqs.(\ref{eq:AB9}) and (\ref{eq:AB7}) we derive: 

\begin{eqnarray}
\frac{\partial \tilde{n}_{k}(t)}{\partial t} && = -\frac{1}{2} [\bra{c} \mathcal{H_I} \ket{v} \tilde{\rho}_{vc}(t) e^{-i \Delta_k t}+ h.c.],
\label{eq:N_undamped}
\end{eqnarray}
and
\begin{equation}
\frac{\partial \tilde{\Pi}_{k}(t)}{\partial t}= \frac{1}{4}\bra{v} \mathcal{H_I} \ket{c} \tilde{n}_{k}(t) e^{i \Delta_k t} .
\label{eq:P_undamped}
\end{equation}
In the above, the `population inversion' $n_k$ is defined as
\begin{equation}
\tilde{n}_k(t)= \tilde{\rho}_{c_k c_k}(t)- \tilde{\rho}_{v_k v_k}(t), \\
\end{equation}
whereas `polarization or inter-band coherence' is given by 
\begin{equation}
\tilde{\Pi}^{*}_{k}(t)=\tilde{\rho}_{c_k v_k}(t) \,\,\, \mathrm{,}\,\,\, \tilde{\Pi}_{k}(t)=\tilde{\rho}_{v_k c_k}(t) .
\end{equation}

As we have taken $ \mathcal{H}_I = (e v_F E/ \omega) \sigma_x$, we obtain the required off-diagonal matrix elements of $\mathcal{H}_I$ ($\bra{c} \mathcal{H_I} \ket{v}$ and $\bra{v} \mathcal{H_I} \ket{c}={\bra{c} \mathcal{H_I} \ket{v}}^{*}$), using the form of conduction and valance band states, ($\ket{c}$ and $\ket{v}$ ) corresponding to the case of graphene, silicene and $\mathrm{MoS_2}$, given by Eqs.~(\ref{eq:gcket}),~(\ref{eq:gvket}),~(\ref{eq:scket}),~(\ref{eq:svket}), ~(\ref{eq:mcket}) and ~(\ref{eq:mvket}).
We find, for graphene:
 \begin{eqnarray}
\bra{v} \mathcal{H}_I \ket{c}&& = (e v_F E/ \omega) \bra{v} \sigma_x \ket{c}\notag \\ &&
 =i\frac{ \, e \, v_F \,E }{ \omega} \sin{\chi_k}\notag \\ &&
 = 2 i \Omega_{vc},
 \end{eqnarray}
 where $\Omega_{vc}=\Omega_k = \frac{e v_F E}{2 \omega} \sin{\chi_k}$, is the so-called `Rabi frequency'. Note that these matrix elements are independent of $s$ and $\tau$ i.e. are same for $s_z=\pm \frac{1}{2}$ and $\tau=\pm 1$ (K and K' valleys).\\
 For silicene : 
 \begin{eqnarray}
\bra{v} \mathcal{H}_I \ket{c}&&= \alpha_{s \tau}\left(\frac{e v_F E}{\omega}\right)(-i \sin{\chi_k} \, |\epsilon_{s \tau}|- \cos{\chi_k} \, \Delta_{s \tau}) \notag \\ &&
= \alpha_{s \tau} (- 2 \,i \, \Omega_k \, |\epsilon_{s \tau}| - 2 \Omega'_k \, \Delta_{s \tau})\notag \\ &&
= 2 i \Omega_{vc},
 \end{eqnarray}
where $ \Omega_{vc}= \alpha_{s \tau} (- \Omega_k \, |\epsilon_{s \tau}| + i \, \Omega'_k \, \Delta_{s \tau})$ and $\Omega_{cv}=\Omega^{*}_{vc}$.\\

For $\mathrm{MoS_2}$:
\begin{eqnarray}
\bra{v} \mathcal{H}_I \ket{c} && = \alpha_{s \tau}\left(\frac{e v_F E}{\omega}\right)(i \sin{\chi_k} \, |\epsilon_{s \tau}|- \cos{\chi_k} \, \Delta_{s \tau}) \notag \\ &&
= \alpha_{s \tau} ( 2 \,i \, \Omega_k \, |\epsilon_{s \tau}| - 2 \Omega'_k \, \Delta_{s \tau})\notag \\ &&
=2 i \Omega_{vc} \,.
\end{eqnarray}
Here $ \Omega_{vc}= \alpha_{s \tau} ( \Omega_k \, |\epsilon_{s \tau}| + i \, \Omega'_k \, \Delta_{s \tau})$ for $\mathrm{MoS_2}$ case. \\
In all three cases
 \begin{equation}
 \bra{c} \mathcal{H}_I \ket{v} ={\bra{v} \mathcal{H}_I \ket{c}}^{*} =-2 i \, {\Omega}^{*}_{vc}.
 \end{equation}
 $\alpha_{s \tau}= \frac{k v_F}{|\epsilon_{s \tau}|\sqrt{|{\epsilon_{s \tau}}|^2 - {\Delta^2_{s \tau}}}}$, $\Omega_k = \frac{e v_F E}{2 \omega} \sin{\chi_k}$ and $\Omega'_k = \frac{e v_F E}{2 \omega} \cos{\chi_k}$. Furthermore, $|\epsilon_{s \tau}|$ and $\Delta_{s \tau}$ have been previously defined in the respective sub-sections on silicene and $\mathrm{MoS_2}$.

\subsection{3b. Dissipative Dynamics}
In this sub-section we introduce the relaxation effects which are present in the system because of its interaction with the surrounding heat bath. These effects account for dissipation that characterizes transition of the system from one equilibrium configuration to another. Suffice it to say, the underlying effects are intrinsic to the system, ever present even in the absence of the external perturbation (the electric field, in the present case) – because of spontaneous statistical fluctuations. Thus, our strategy is to examine the dissipative phenomenon in the absence of the external field, by expanding the unperturbed Hamiltonian to a form that incorporates the system-plus-bath coupling. Once we have a master equation for the density operator at hand (with dissipative or relaxational terms present) we will add that to the contribution arising from the systematic component, as dealt with in  section 3a. \\  
The system-plus-bath Hamiltonian will have to account for two distinct physical interactions between the Dirac electron and the surrounding phonons and other electrons. The phonons that describe lattice relaxations can cause transitions between the conduction and valence bands thus maintaining the system in thermal equilibrium governed by the Fermi-Dirac distribution. Thus these transitions involve `depopulation' - a terminology well known in vibrational relaxation in molecular spectra~\cite{dattagupta} -- of the valence and conduction levels. Evidently, the coupling to the phonons will have to occur via a term that is purely off-diagonal in the basis states of conduction and valence bands, thus yielding:
\begin{equation}
\mathcal{H}_{\mathrm{el-ph}}=(\ket{c}\bra{v} + \ket{v}\bra{c} ) \sum_{q} g_{q} (b_q + b^{\dagger}_q) + \sum_{q}  \Omega_q b^{\dagger}_q b_q \,.
\end{equation}
Here, $g_q$ is a coupling constant and the last term denotes a `free' phonon Hamiltonian, the $b$’s and $b^{\dagger}$‘s being the annihilation and creation operators for phonons. \\
On the other hand, there exists a much stronger interaction with the other electrons which however do not trigger any transition between the two bands but simply cause `de-phasing' thereby leading to what is known as `dissipation-less de-coherence' ~\cite{SDSP}  . The corresponding interaction can be represented by
\begin{equation}
\mathcal{H}_{el-el}=(\ket{c}\bra{c} - \ket{v}\bra{v})\sum_{p} G_p (c_p + c^{\dagger}_p) + \sum_{p}  \omega_p c^{\dagger}_p c_p \,.
\end{equation}
Once again, $G_p$   parametrizes the coupling with the electrons. Evidently, the coupling term now commutes with the unperturbed Hamiltonian and hence, does not influence any energy exchange between the system and the surrounding electronic bath. The last term contains the free electron interaction in terms of electron creation and annihilation operators. The two level system in coupling with a fermionic bath has been treated in detail in~\cite{chang}. As shown there, the bath(fermionic) can be `bosonized' as long as we are only interested in the electron-hole excitations near the fermi surface, for what is called `Ohmic dissipation'. We will adopt that interpretation here, since we are only considering low energy Hamiltonian and excitations near the K/K' valley points.\\
The full system-plus-bath Hamiltonian can then be written as
\begin{eqnarray} 
\mathcal{H}_{SB}&&= \epsilon_c \ket{c}\bra{c}  + \epsilon_v \ket{v}\bra{v}\notag \\ &&+ (\ket{c}\bra{v} + \ket{v}\bra{c} ) \sum_{q} g_{q} (b_q + b^{\dagger}_q) + \sum_{q}  \Omega_q b^{\dagger}_q b_q \notag \\ && + (\ket{c}\bra{c} - \ket{v}\bra{v})\sum_{p} G_p (c_p + c^{\dagger}_p) \notag \\ && + \sum_{p}  \omega_p c^{\dagger}_p c_p \,.
\label{eq:HSB}
\end{eqnarray}
The Hamiltonian given above in Eq.~(\ref{eq:HSB}) can form the basis of a detailed study of quantum dissipative phenomena for the two-band materials at hand, following the methods reviewed in~\cite{weiss11}. While relegating that investigation to the future we study here a simpler model in which the phonon and electronic bath operators are replaced by effective fields and follow standard treatments of master equations (chapter 1) of ~\cite{SDSP}. We should note however that because of the harmonic nature of the phonon and electron baths and the assumed linear coupling, as is prevalent also in Feynman-Vernon, Caldeira-Leggett  treatments (See~\cite{weiss11}), only Gaussian fluctuations are relevant. Thus, Cumulant Expansion methods, as detailed in Appendix I.A.I of ~\cite{SDSP} , are the most suitable because all cumulants beyond the second order vanish.\\
 The simplified system-bath Hamiltonian can then be rewritten as
\begin{eqnarray} 
\mathcal{H}_{SB}&&= \epsilon_c \ket{c}\bra{c} + \epsilon_v \ket{v}\bra{v}+ g (\ket{c}\bra{v} + \ket{v}\bra{c} ) \textbf{B} \notag \\ &&  + \mathcal{H}_{ph}  + G (\ket{c}\bra{c} - \ket{v}\bra{v})\textbf{C}  + \mathcal{H}_{el} \, .
\label{eq:HSB1}
\end{eqnarray}
In the above, g and G are the phonon and electron coupling constants (in which their momentum-dependence has been ignored), $\textbf{B}$ and $\textbf{C}$ are the classical version of phonon and electron `displacement operators', and $\mathcal{H}_{ph}$ and $\mathcal{H}_{el}$  are the free phonon and electron Hamiltonians, respectively.\\
Following the results given in  I.7 of~\cite{SDSP}, the different components of the dissipative part of the density operator, in the interaction picture of the unperturbed Hamiltonian $\mathcal{H}_0$, can be shown to satisfy 
\begin{equation}
\frac{\partial }{\partial t} \bra{v} \tilde{\rho}_{D} \ket{v}= \lambda (p_v \bra{c} \tilde{\rho}_D \ket{c}-p_c \bra{v} \tilde{\rho}_D \ket{v}), 
\label{eq:p1}
\end{equation}
\begin{equation}
\frac{\partial }{\partial t} \bra{c} \tilde{\rho}_{D} \ket{c}= -\lambda (p_v \bra{c} \tilde{\rho}_D \ket{c}-p_c \bra{v} \tilde{\rho}_D \ket{v}),
\label{eq:p2}
\end{equation}
\begin{equation}
\frac{\partial }{\partial t} \bra{c} \tilde{\rho}_{D} \ket{v}= \frac{(\lambda + \gamma)}{2} [ \bra{v} \tilde{\rho}_D \ket{c}- \bra{c} \tilde{\rho}_D \ket{v}],
\end{equation}
\begin{equation}
\frac{\partial }{\partial t} \bra{v} \tilde{\rho}_{D} \ket{c}= \frac{(\lambda + \gamma)}{2} [ \bra{c} \tilde{\rho}_D \ket{v}- \bra{v} \tilde{\rho}_D \ket{c}].
\end{equation}
The Eq.~(\ref{eq:p2}) is consistent with the conservation of probability: $ \bra{c} \tilde{\rho}_{D} \ket{c}+ \bra{v} \tilde{\rho}_{D} \ket{v}=1 $. 
In the following we treat $\lambda$ and $\gamma$ as lumped parameters which have been listed by~\cite{couplings} as $\lambda \approx 10^{12} $ Hz while $\lambda+\gamma=\Upsilon \approx 10^{14}$ Hz for graphene. Here, we identify that, $\Upsilon=\lambda+\gamma$ is the total rate due to the combined phonon and electron bath.\\
It turns out that~\cite{SDSP}, for phonon bath,
\begin{equation}
\lambda\, p_{v,c}=g^2 \int_{-\infty}^{\infty} d\tau <\textbf{B}(\tau)\textbf{B}(0)>e^{\pm i \omega_{cv}t},
\end{equation}
where plus/minus sign in the exponential is for valence/conduction band index and $<\textbf{B}(\tau)\textbf{B}(0)>$ is an auto-correlation of the phonon bath operators defined in the interaction picture as
\begin{equation}
\textbf{B}(\tau)=e^{i \mathcal{H}_{ph}t} \textbf{B}(0) e^{-i\mathcal{H}_{ph} t } \,.
\end{equation}
Note that the relation between the Fermi distribution $p_{c}$ and $p_{v}$(detailed balance) follows from the spectral properties of the correlation function. The equilibrium population is then,  $ n^{eq}_k=p_{c}-p_{v} $.
 For electron bath, the relaxation rate $\gamma$, on the other hand, is given by 
 \begin{equation}
 \gamma=G^2 \int_{-\infty}^{\infty} d\tau <\textbf{C}(\tau)\textbf{C}(0)>\, ,
\end{equation}   
where $<\textbf{C}(\tau)\textbf{C}(0)>$ is an auto-correlation of the electron bath operators defined in the interaction picture as,
\begin{equation}
\textbf{C}(\tau)=e^{i \mathcal{H}_{el}t} \textbf{C}(0) e^{-i\mathcal{H}_{el} t }.
\end{equation}
 The coupling constants $g$ and $G$ (on which the rates $\lambda$ and $\gamma$ depend) can be calculated from microscopic considerations of electron-phonon scattering and electron-electron scattering for different materials e.g.  graphene as discussed in~\cite{grapheneEP1,grapheneEP2,grapheneEP3,grapheneEP4} and~\cite{grapheneEE0,grapheneEE1,grapheneEE2,grapheneEE3}, silicene~\cite{SiliceneEP1,SiliceneEP2} and $\mathrm{MoS}_2$~\cite{MoS2EP1,MoS2EP2}.\\ 
In the above analysis we have obtained the dissipative evolution part and  the systematic evolution part of the Eq.~(\ref{eq:RhoEvoltuion}) as given by the Eq.~(\ref{eq:N_undamped}) and Eq.~(\ref{eq:P_undamped}), where the matrix elements  $\bra{v} \mathcal{H_I} \ket{c}= 2 i \Omega_{vc}$ and $\bra{c} \mathcal{H_I} \ket{v}= -2 i \Omega_{cv}$ and expressions for $\Omega_{vc}$ (and $\Omega_{cv}=\Omega^{*}_{vc}$) have been stated  in the previous section corresponding to our three cases. We are now ready to write down the  equation~(\ref{eq:RhoEvoltuion}) in component form (in the interaction picture), thereby yielding 
\begin{eqnarray}
\frac{\partial \tilde{n}_{k}(t)}{\partial t} &&= i [\Omega_{cv}  \tilde{\Pi}_{k}(t) e^{-i \Delta_k t}
- \Omega_{vc}   \tilde{\Pi}^{*}_{k}(t) e^{i \Delta_k t}] \notag \\ &&
- \lambda \, (\tilde{n}_k (t) - n^{eq}_k ) ,
\end{eqnarray}

\begin{eqnarray}
\frac{\partial \tilde{\Pi}_{k}(t)}{\partial t} &&= i \frac{\Omega_{vc}}{2} \tilde{n}_{k}(t) e^{i \Delta_k t}  \notag \\ &&
-\Upsilon \,\tilde{\Pi}_{k}(t).
\end{eqnarray}

Let us define a following transformation for the elements of the density matrix, 
\begin{equation}
\begin{split}
\Pi'_{k}(t)= \tilde{\Pi}_{k} \, e^{-i \Delta_k t} \,\, , \,\, \Pi'^{*}_{k}(t)= \tilde{\Pi}^{*}_{k} \, e^{i \Delta_k t}\\
\rho'_{cc}= \tilde{\rho}_{cc} \,\,, \,\, \rho'_{vv}= \tilde{\rho}_{vv} \rightarrow n'_k (t)= \tilde{n}_k (t)\, . \label{eq:T2}
\end{split}.
\end{equation}
Then the above equations becomes
\begin{equation}
\frac{\partial n'_{k}(t)}{\partial t}= i [\Omega_{cv}  \Pi'_{k}(t)
- \Omega_{vc}   \Pi'^{*}_{k}(t) ]
- \lambda \, (\tilde{n}_k (t) - n^{eq}_k ) ,
\end{equation}

\begin{equation}
\frac{\partial \Pi'_{k}(t)}{\partial t}= -i \Delta_k \, \Pi'_{k}(t) + i \frac{\Omega_{vc}}{2} n'_{k}(t) 
-\Upsilon \,\Pi'_{k}(t).
\end{equation}
The steady state solutions for these equations are given by
\begin{equation}
n'_k= \lambda \, n^{eq}_k \left( \frac{\Upsilon^2 + \Delta^{2}_k}{ \lambda \,\Upsilon^2 + \lambda \, \Delta^{2}_k + \Upsilon \, |\Omega_{vc}|^2 }   \right),
\label{eq:solN}
\end{equation}

\begin{equation}
\Pi'_{k}=\frac{n'_k \, \Omega_{vc} }{2 ( \Delta_k - i \, \Upsilon )}
\label{eq:solP1}.
\end{equation}

\section{4. Nonlinear Optical Conductivity}

For calculating the conductivity we need the cycle-averaged power absorbed, defined by 

\begin{equation}
(\vec{J}(t).\vec{E}(t))_{av}= \frac{\omega}{2 \pi} \int_{0}^{2 \pi/{\omega}} \vec{J}(t).\vec{E}(t) \label{eq:1cycle} \, .
\end{equation}
The general expression for the d-dimensional system is given by
\begin{equation}
\vec{J}(t)= - \frac{g_s g_v}{2 {\pi}^{d}} \int d \vec{k} \, {\vec{J}}_{\vec{k}} (t) \, \, \label{eq:J} , 
\end{equation}
where $g_s\, ( g_v)$ is the spin (valley) degeneracy factor(in our case both are 2) and momentum dependent component of particle current density
\begin{equation}
\vec{J}_{\vec{k}}(t)= e \, Tr[\rho_{\vec{k}}(t) \vec{v}_{\vec{k}}(t)],
\end{equation}
and $x$-direction component of this,
\begin{equation}
\vec{J}_{\vec{k}x}(t)= e \, Tr[\rho_{\vec{k}}(t) v_{\vec{k}x}] = e \, v_F \, Tr[\rho_{\vec{k}}(t) \, \sigma_x]\, .
\end{equation}
A general expression for the $\vec{J}_{\vec{k}x}(t)$ can be calculated, given the Hamiltonian and the valence and conduction eigenstates for any 2-band system, which in our case
\begin{eqnarray}
\textbf{J}_{\vec{k}x}(t)&=& e   v_F  \frac{\tau   k  v_F}{|\epsilon_{s\tau}|} \cos{\chi_k}   ~ n(t)\nonumber\\
&+&   e  v_F  \sin{\chi_k} \left[i\frac{\Omega_{vc}}{\Omega_k}\Pi^{*}_{k}(t)+ h.c.\right] .
\label{eq:Jx}
\end{eqnarray}
In order to convert this expression in terms of transformed density matrix's elements ($\Pi'_{k},\, \Pi'^{*}_{k},\, n'_k$) we use Eqs.~(\ref{eq:T1}) and ~(\ref{eq:T2}) to  get $\Pi_{k}(t)= \Pi'_{k} \, e^{i \omega t}$, $\Pi^{*}_{k}(t)= \Pi'^{*}_{k} \, e^{-i \omega t}$ and $n_k(t)=n'_k$. When we substitute Eq.~(\ref{eq:Jx}) in to the Eq.~(\ref{eq:1cycle}) and  integrate over a complete cycle of the applied field, the first term of Eq.~(\ref{eq:Jx}) yields zero while the second term  gives non-zero contribution. Therefore we will consider  from  Eq.~(\ref{eq:Jx}) only  the second term, i.e., 
\begin{eqnarray}
\textbf{J}_{\vec{k}x}(t)&=&   \,e \, v_F \, \sin{\chi_k} \,\,\left[i\,\frac{\Omega_{vc}}{\Omega_k}\Pi'^{*}_{k} e^{-i \omega t}+ h.c. \right] . \label{eq:Jx_Eff}
\end{eqnarray}
Using the solution for $\Pi'_{k}$ from Eq.~(\ref{eq:solP1})  in Eq.~(\ref{eq:Jx_Eff}) yields,
\begin{equation}
\textbf{J}_{\vec{k}x}(t) = n'_k  \frac{(e v_F  \sin{\chi_k} |\Omega_{vc}|^{2})}{\Omega_k} \frac{(\Upsilon \cos{\omega t - \Delta_k \sin{\omega t}})}{(\Delta^{2}_{k} + \Upsilon^2)},
\end{equation}
where $\Omega_{cv}$ and $\Omega^*_{vc}=\Omega_{cv}$ have been given in the previous sections corresponding to the  three cases, while   $n'_k(t)$ is given by Eq.~(\ref{eq:solN}) and $\Delta_k=\omega - \omega_{cv}$. It should be noticed that only the term proportional to $\cos{\omega t}$ (out of phase with the applied term of Eq.~(\ref{eq:I7}) ) is responsible for dissipation that contributes to the conductivity and the other term proportional to $\sin{\omega t}$ will be dropped from further consideration. Therefore,
\begin{equation}
\textbf{J}_{\vec{k}x}(t) = n'_k  \frac{(e\, v_F \, \sin{\chi_k} \, |\Omega_{vc}|^{2})}{\Omega_k} \frac{\Upsilon \cos{\omega t}}{(\Delta^{2}_{k} + \Upsilon^2)}. \label{eq:current}
\end{equation}
The general expression for the longitudinal non-linear optical conductivity is given by,
\begin{equation}
\sigma_{xx}(\omega)= -\frac{g_s g_v}{(2 \pi)^2}\int d \vec{k} \,\,\frac{|\bar{\textbf{J}}_{\vec{k}x}(t)|}{E_0 \, \cos{\omega t}}. \label{eq:Conductivity}
\end{equation}
We next write down the expressions for $\textbf{J}_{\vec{k}x}(t)$ for  the three cases of interest. 
For graphene, we have $\Omega_{vc}=\Omega_k=\frac{e v_F E_0 \sin{\chi_k}}{2 \omega}=\Omega_{cv}$, $\Delta_k = \omega - \omega_{cv}$ and $\omega_{cv}=2kv_F$, plugging these into the Eq.~(\ref{eq:current}) one gets,
\begin{equation}
\textbf{J}_{\vec{k}x}(t) = n'_k  \frac{(e v_F \sin{\chi_k})^2}{2 \omega} \frac{\Upsilon }{(\Delta^{2}_{k} + \Upsilon^2)}(E_0 \,\cos{\omega t}) \label{eq:Jgraphene}
\end{equation}

For silicene, we have $\Omega_{cv}=\alpha_{s \tau}(-\Omega_k |\epsilon_{s \tau}|+ i \Omega'_{k} \Delta_{s\tau})$, $\Omega^{*}_{vc}=\Omega_{cv}$, $\Omega_{cv}=2 |\epsilon_{s \tau}|=2\sqrt{k^2 v^2_F+\Delta^2_{s \tau}} $ , $\Omega_k=\frac{e v_F E_0 \sin{\chi_k}}{2 \omega}$, $\Omega'_k=\frac{e v_F E_0 \cos{\chi_k}}{2 \omega}$ and $\alpha_{s \tau}=\frac{k v_F}{|\epsilon_{s \tau}| \sqrt{|\epsilon_{s \tau}|^2-\Delta^2_{s \tau}}}$. putting these in the Eq.~(\ref{eq:current}) gives us,
\begin{equation}
\begin{split}
\textbf{J}_{\vec{k}x}(t) = n'_k \frac{(e\, v_F\, \alpha_{s \tau})^2 (\sin^2{\chi_k} |\epsilon_{s \tau}|^2 + \cos^2{\chi_k} \Delta^2_{s \tau} )}{2 \omega}\\
\times \frac{\Upsilon }{(\Delta^{2}_{k} + \Upsilon^2)}(E_0 \,\cos{\omega t}) .
\end{split}\label{eq:JSilicene}
\end{equation}
Finally, for $\mathrm{MoS_2}$  we have $\Omega_{cv}=\alpha_{s \tau}(\Omega_k |\epsilon_{s \tau}|+ i \Omega'_{k} \Delta_{s\tau})$ and $\Omega^{*}_{vc}=\Omega_{cv}$. Plugging it in  Eq.~(\ref{eq:current}) yields a term identical to  Eq.~(\ref{eq:JSilicene}).
Thus  the longitudinal optical conductivity in all  the three cases can be calculated by plugging the corresponding $\textbf{J}_{\vec{k}x}(t)$ in Eq.~(\ref{eq:Conductivity}) and integrating over the allowed  range of momentums (near  the vicinity of K/K' points) for which the Dirac Hamiltonian form is valid.

\section{5. Results and Analysis}
 Employing $\Omega_{vc}=\frac{\tau e v_F E_0}{\omega} \sigma^x_{vc}$, then Eq.~(\ref{eq:current}) can be cast in the following form
\begin{equation}
\textbf{J}_{\vec{k}x}(t) = -n'_k \frac{e^2 v^2_{F}|\sigma^x_{vc}|^2}{\omega} \frac{ \Upsilon}{(\omega_k-\omega)^2 + \Upsilon^2}(E_0 \cos(\omega\, t)),
\label{eq:Jnew}
\end{equation}
 where,
\begin{equation}
|\sigma^x_{vc}|^2= \frac{\Delta^2_{s\tau}+ k^2 v^2_F \sin^2(\chi_k)}{\Delta^2_{s\tau}+k^2 v^2_F}
\label{eq:Sigma2}.
\end{equation} 
  And $n'_k $ is given by the Eq.~(\ref{eq:solN}), which can also be written in the following form after defining dimensionless parameter $\xi=\frac{e v_F E_0}{\omega \sqrt{\lambda \Upsilon}}$ characterizing linear regime ($\xi \ll1 $) and non-linear regime($\xi\geq 1$),
\begin{equation}
n'_k = n^{eq}_k \left(1 + \xi^2 |\sigma^x_{vc}|^2 \frac{\Upsilon^2}{[(\omega_k-\omega)^2+\Upsilon^2)]} \right)^{-1}
\label{eq:Nnew}.
\end{equation} 
In our cases, spin and valley degeneracy factors $g_s,g_v=2$ and putting the expression for $\textbf{J}_{\vec{k}x}(t)$ in Eq.~(\ref{eq:Conductivity}) from Eq.~(\ref{eq:Jnew}), we get,
\begin{equation}
\begin{split}
\sigma_{xx}(\omega)=\frac{1}{\pi^2}\int^{2 \pi}_{0}d\chi_k \int k \, dk \, n'_k \frac{e^2 v^2_{F}|\sigma^x_{vc}|^2}{\omega}\\ \times \frac{ \Upsilon}{(\omega_k-\omega)^2 + \Upsilon^2}
\end{split}
\label{eq:Conductivity2}
\end{equation}
where $|\sigma^x_{vc}|^2$ and $n'_k$ are respectively given by Eq.~(\ref{eq:Sigma2}) and Eq.~(\ref{eq:Nnew}).
Note that the Eq.~(\ref{eq:Conductivity2}) is a simplified version of  Eq.~(22) of ~\cite{agarwal}. The latter authors in~\cite{agarwal} have made further discussion of the longitudinal optical conductivity in detail for different cases-linear and non-linear in different regimes-clean and dirty. We can derive results using Eq.~(\ref{eq:Conductivity2}) for silicene and $\mathrm{MoS}_2$ which however appear to match those of Agarwal et al. for `gapped' graphene. We do not repeat that analysis here.

\section{6. Concluding Remarks}
We have calculated the non-linear optical conductivity for gap-less(graphene) and gapped (silicene and $\mathrm{MoS_2}$) two band systems. We follow a density matrix approach to study the time evolution of these systems. The treatment presented in section 3 is the core of the paper that offers a relaxation model for describing dissipative response of the 2-D materials at hand, to an externally applied frequency dependent field. After obtaining the time evolution equations for different components of the density matrix including damping terms, we solve for the case of steady state and get analytical expressions for population inversion and inter-band coherence given by Eq.~(\ref{eq:solN}) and Eq.~(\ref{eq:solP1}). A comment is in order on clarifying the difference in our treatment from that of~\cite{mishchenko}. The phenomenological equations(10) of ~\cite{mishchenko} incorporate a damping parameter $\Gamma_p$ in however the wave function equations. The problem with this approach is that the total probability, in the presence of dissipation, is not conserved. Our method circumvents this issue by looking at a fully quantum master equation for density operator, besides taking care of two time-scales (inverse relaxation rates) associated with electron-phonon and electron-electron interactions. \\
As discussed in~\cite{agarwal}, we may categorize the four regimes depending on two parameters $\xi$ (defines non-linearity) and ${\Upsilon}/{\omega}$ (characterizes clean or dirty system) as follow: (a) linear clean limit (in which $\xi \ll 1$ and $\Upsilon / \omega \ll1$ ), (b) linear dirty limit (in which $\xi\ll1$ and $\Upsilon/ \omega \geq 1$ ), (c) non-linear clean limit (in which $\xi \geq 1$) and $\Upsilon / \omega \ll1$ 
 and (d) non-linear dirty limit (in which $\xi \geq 1$ and $\Upsilon / \omega \geq 1$). The parameter $\xi$ first appears naturally in the steady state solutions for population inversion and inter-band coherence and then in the conductivity expressions in various cases. If we assume $\xi\ll1 $, the results match the linear response results, that's why $\xi$ specifies the strength of optical non-linearity in the system. Furthermore, for $\Upsilon/ \omega \ll1$, the term $\frac{\Upsilon}{(\omega_k - \omega)^2 + \Upsilon^2} \rightarrow \delta(\omega_k - \omega)$ in the momentum dependent conductivity expression (Eq.~(\ref{eq:Jnew})), which means this non-zero contribution comes only from the frequency $\omega = \omega_k $, that's why $\Upsilon / \omega \ll1$ is called `clean limit'. When $\Upsilon/ \omega \geq 1 $, then nearby frequencies to $\omega=\omega_k$ can also contribute to the optical conductivity and hence is called `dirty limit'.  \\
 Finally, with the steady state solutions for density matrix components in hand, we can calculate the optical conductivity in these four regimes and obtain analytical solutions for most of the cases in $T \rightarrow 0$ limit. Some specific cases need numerical evaluation for momentum dependent conductivity integrals.
One may note that the linear response theory (based on the Kubo formula)  for conductivity is only valid in the high frequency limit ($\Upsilon / \omega \ll 1 $). \\
In conclusion, our ability to derive the known results for non-linear optical conductivity in gapless and gapped graphene, based on a simplified treatment of the Hamiltonian  Eq.~(\ref{eq:HSB}), may spur further and more extensive study of what is generally known as Spin-Boson Hamiltonian (Eq.~(\ref{eq:HSB})) in the context of quantum dissipative systems~\cite{weiss11}.  \\
 
\section{7. Acknowledgement}
S.D. is grateful to the Indian National Science Academy for support through their Senior Sciencist scheme and to IISER, Bhopal for kind hospitality. We thank Suhas Gangadharaiah for useful discussions in the early stages of the work.

\end{document}